\pdfoutput=1
\documentclass[12pt]{article}
\usepackage{graphicx}
\usepackage{xspace}
\usepackage{paralist}
\usepackage{hyperref}
\usepackage{fancyvrb}
\usepackage[usenames,dvipsnames,table]{xcolor}
\usepackage{amsmath}

\newcommand{\rep}[1]{\ensuremath{\boldsymbol{#1}}}
\newcommand{\crep}[1]{\ensuremath\overline{\boldsymbol{#1}}}
\newcommand\pubnumber{DPF2015-225}
\newcommand\pubdate{\today}
\newcommand{\code}[1]{{{\tt #1}}}

\newcommand{\pyrate}{PyR@TE\xspace}
\newcommand{\pyratetwo}{PyR@TE\xspace2\xspace}
\newcommand{\python}{\code{Python}\xspace}
\newcommand{\sarah}{\code{SARAH}\xspace}
\newcommand{\mathematica}{\code{Mathematica}\xspace}
\def\CC{{C\nolinebreak[4]\hspace{-.05em}\raisebox{.4ex}{\tiny\bf ++}}\xspace}

\def\smu{Southern Methodist University\\
	Dallas,Texas,75206, USA}

\def\Title#1{\begin{center} {\Large #1 } \end{center}}
\def\Author#1{\begin{center}{ \sc #1} \end{center}}
\def\Address#1{\begin{center}{ \it #1} \end{center}}

\newcommand\pubblock{\rightline{\begin{tabular}{l} \pubnumber\\
         \pubdate  \end{tabular}}}
\newenvironment{Abstract}{\begin{quotation}  }{\end{quotation}}
\newenvironment{Presented}{\begin{quotation} \begin{center} 
             PRESENTED AT\end{center}\bigskip 
      \begin{center}\begin{large}}{\end{large}\end{center} \end{quotation}}
\def\Acknowledgments{\bigskip  \bigskip \begin{center} \begin{large}
             \bf ACKNOWLEDGMENTS \end{large}\end{center}}

\def\beq{\begin{equation}}
\def\eeq#1{\label{#1}\end{equation}}
\def\eeqn{\end{equation}}

\def\beqa{\begin{eqnarray}}
\def\eeqa#1{\label{#1}\end{eqnarray}}
\def\eeqan{\end{eqnarray}}

\let\bar=\overbar

\def\Dslash{\not{\hbox{\kern-4pt $D$}}}
\def\dslash{\not{\hbox{\kern-2pt $\del$}}}

\def\msb{{\bar{\ssstyle M \kern -1pt S}}}

\begin{document}
\begin{titlepage}
\pubblock

\vfill
\Title{Automation of non-SUSY two-loop RGEs with PyR@TE: latest developments}
\vfill
\Author{Florian Lyonnet}
\Address{\smu}
\vfill
\begin{Abstract}
	In light of the conspicuous absence of SUSY in the energy range explored by the LHC during run I, non-supersymmetric BSM scenarios are becoming more and more attractive. One key ingredient in exploring such BSM physics are the renormalization group equations (RGEs) that are essential for extrapolating the theory to higher energy scales. Although the two-loop RGEs for a general quantum field theory have been known for some time, it is only recently that their automation has become available in the form of a Python program called PyR@TE. In this talk, I will present the features of PyR@TE as well as the latest developments of the code. In particular, the new ability to deal with sets of fields that have multiple ways of being contracted into a gauge singlet.
\end{Abstract}
\vfill
\begin{Presented}
DPF 2015\\
The Meeting of the American Physical Society\\
Division of Particles and Fields\\
Ann Arbor, Michigan, August 4--8, 2015\\
\end{Presented}
\vfill
\end{titlepage}
\def\thefootnote{\fnsymbol{footnote}}
\setcounter{footnote}{0}

\section{Introduction}

The second run of the LHC is now ongoing and new data will soon allow us to shed more light on the physics living at the TeV scale. Even though SUSY is one of the best motivated candidates for this new physics, the absence of evidence for it so far has revived the interest in non-supersymmetic (SUSY) extensions of the Standard Model (SM). 

In the SUSY context, there exist two \mathematica packages that allow for an automatic generation of all the $\beta$-functions at two-loop, \sarah ~\cite{Staub:2013tta} and SUSYNO~\cite{Fonseca:2011sy}. \pyrate~\cite{Lyonnet:2013dna} bridges the gap with SUSY models and brings to the same level of automation the calculation of two-loop RGEs for non-SUSY extensions of the SM.

In this presentation, we review the capabilities of \pyrate and present the latest developments of the code. In particular, we show how to deal with terms that have multiple gauge singlets. Finally, we provide an outlook on the upcoming release of \pyratetwo~\cite{lyonnet:2016pyr}. 

\section{Renormalization group equations at two-loop with \pyrate}

\pyrate~\cite{Lyonnet:2013dna} is a \python program that automatically generates the full two-loop renormalization group equations for all the (dimensionless and dimensionful) parameters of a general gauge theory . The general RGEs for non-supersymmetic gauge theories have been known at two-loop accuracy for about 30 years~\cite{Machacek:1983tz,Machacek:1983fi,Machacek:1984zw,Jack:1982hf,Jack:1982sr,Jack:1984vj}. All known typos in the original series of papers by Machacek and Vaughn have been taken into account and the code has been extensively validated against the literature. In addition, independently of \pyrate, \mathematica routines were developed and cross-checked against \pyrate; these routines are now part of \sarah 4~\cite{Staub:2013tta}. 

The various inputs such as the gauge group and particle content can be specified by the user via simple text files. For instance, the Higgs quartic term of the SM Lagrangian would be defined by

\begin{Verbatim}[numbers=left,xleftmargin=20pt,formatcom=\color{cyan}]
QuarticTerms: {
	'\lambda': {Fields : [H,H*,H,H*], Norm : 1/2}}
  }
\end{Verbatim}
In which \verb|H|, \verb|H*| represent the Higgs field and its conjugate respectively.

Once the RGEs have been calculated by \pyrate, the results can be exported to \LaTeX and \mathematica or stored in a \python data structure for further processing. In addition, it is possible to create a numerical \python function that can be used to solve the RGEs, e.g. using the provided \python tool box\footnote{Note that a \mathematica routine to solve the exported RGEs is also provided.}.

In terms of gauge groups, \pyrate currently supports $\mathrm{SU}(n)$ groups\footnote{Higher $\mathrm{SU}(n)$ gauge groups can be added upon request.} up to $\mathrm{SU}(6)$ as well as $\mathrm{U}(1)$. For each one of these groups, the irreducible representations available range from dimension $n$ to $n^2-1$ with the exception of $\mathrm{SU}(2)$ and $\mathrm{SU}(3)$ for which an extended database is available with representations up to dimension $8$ and $10$ respectively. Note that the kinetic mixing between multiple $\mathrm{U}(1)$ group factors is not yet taken into account, and we refer the reader to section~\ref{sec:outlook} for more details.

The code was designed to provide the user with maximum flexibility with regard to what to calculate. For example, one can select to calculate the $\beta$-function of only one of the terms defined in the potential or even to neglect a specific contribution to a particular RGE. The following command line would calculate the RGEs of the gauge couplings, \verb|-gc|, and quartic term, \verb|-qc|, of the SM at one-loop neglecting the contribution $A_{abcd}$\footnote{We refer the reader to~\cite{Lyonnet:2013dna} for the definition.}, \verb|--Skip/-sk ['CAbcd']|,

\begin{Verbatim}[numbers=left,xleftmargin=20pt,formatcom=\color{cyan}]
python pyR@TE.py -m SM.model -gc -qc -sk ['CAabcd'] 
\end{Verbatim}

These options are particularly useful for quick validations when developing a model or to avoid calculating terms that are time consuming and that will be neglected in the end. The full list of options can be found in Table 1 of~\cite{Lyonnet:2013dna}.

The code and various tutorials are publicly available at \url{http://pyrate.hepforge.org}.

\section{Latest Developments}

The latest version of \pyrate available is 1.2.1. It contains all the bug fixes introduced since the version 1.0.0 as well as a couple of new flags to steer the calculation such as for instance the \verb|-sk/--Skip| flag described above. In addition, we recently added a significant new capability that we now describe.

The main motivation is to be able to enter terms that have multiple gauge singlets. Indeed, as seen above, each term in the Lagrangian is specified by a set of fields making it impossible to distinguish multiple singlets that could result from the contraction of the same set of fields. So far, the chosen combination was the one in which the quartic term can be factorized in 
\begin{equation}
	\underbrace{(\rep{a}\otimes\rep{b})}_{\rep{1}}\otimes\underbrace{(\rep{c}\otimes\rep{d})}_{\rep{1}}\, ,
\end{equation}
in which $\rep{a},\ \rep{b},\ \rep{c},\ \rep{d}$ are the representations of the scalar fields involved. In general, assuming a set of $n$ fields  $\phi_i$ with dimensions $D_i$ under an $\mathrm{SU}(n)$ gauge group, we will denote the Clebsch-Gordan coefficients (CGCs) that gives the contraction of indices to an invariant combination as $\mathcal{C}$, i.e.
\begin{equation}
	 {\cal C}_{i_1 i_2 \dots i_n}   \phi_{i_1} \phi_{i_2} \dots \phi_{i_n}\, ,
 \end{equation}
 does not transform under $\mathrm{SU}(n)$. Here, the $i_x$, $x=1,\ldots,n$ are the charge indices with respect to the gauge group. As an example let us consider a complex triplet of $\mathrm{SU}(2)$ written as a two-by-two matrix $\Delta$; two quartic invariants are usually used in the literature 
\begin{eqnarray}
	&\mathrm{Tr}(\Delta^{\dagger}\Delta\Delta^{\dagger}\Delta)\, ,\label{eq:exsu2_1}\\
	&\mathrm{Tr}(\Delta^{\dagger}\Delta)\mathrm{Tr}(\Delta^{\dagger}\Delta)\, .\label{eq:exsu2_2}
\end{eqnarray}
Note that because a linear combination of CGCs is a valid singlet, there is an arbitrary choice to make in the basis of CGCs.

Our implementation can be decomposed in three steps:\begin{inparaenum}[(i)]\item Extend the database\footnote{The CGCs are taken from SUSYNO~\cite{Fonseca:2011sy}, see \cite{Lyonnet:2013dna} for more details.} to include all the CGCs, \item make the database available to the user such that one can match one's own definitions to the ones in \pyrate, \item introduce a keyword, \verb|CGCs|, to specify which one of the CGCs to use for a given term\end{inparaenum}.  

The database of \pyrate is built on SUSYNO~\cite{Fonseca:2011sy}, and it is natural to solve (i) in the same context. The meaning of (iii) will be apparent when we discuss the example below and we now focus our discussion on (ii).

\subsection*{Querying the database}
To let the user access the different group related information built in \pyrate, we introduce a new mode to {\it interactively query the database}. This mode is started by specifying the \verb|-idb| flag on the command line; the database is loaded and a prompt appear in which one can type any of the following command to access the corresponding information contained in the database:
\begin{itemize}
	\item \verb|Invariants| {\it group irrep} $\Rightarrow$ CGCs of {\it irrep} of {\it group} ,
	\item \verb|Matrices| {\it group irrep} $\Rightarrow$ matrix representation of {\it irrep} of {\it group},
	\item \verb|Casimir| {\it group irrep} $\Rightarrow$ quadratic casimir of {\it irrep} of {\it group},
	\item \verb|Dynkin| {\it group irrep} $\Rightarrow$ dynkin index of {\it irrep} of {\it group},
	\item \verb|DimR| {\it group irrep} $\Rightarrow$ dimension of {\it irrep} of {\it group},
\end{itemize}

%
%
\subsection*{A toy model: SM+complex triplet}

As an example, let us consider the SM extended by a $\mathrm{SU}(2)_L$-complex triplet, $T\sim(3,1)$. Writing the complex triplet as a two-by-two matrix, $\Delta$, one can write the following potential
\begin{eqnarray}
	\mathcal{V} =& \lambda_1 H^\dagger H H^{\dagger} H + \lambda_{\Delta_1}\mathrm{Tr}(\Delta^{\dagger}\Delta)H^{\dagger}H
	+\lambda_{\Delta_2}\mathrm{Tr}(\Delta^{\dagger}\Delta)\mathrm{Tr}(\Delta^{\dagger}\Delta)\nonumber\\
	&+\lambda_2 H^{\dagger}\Delta \Delta^{\dagger}H + \lambda_3 \mathrm{Tr}(\Delta^{\dagger}\Delta\Delta^{\dagger}\Delta)\, ,
\end{eqnarray}
in which $H$ denotes the Higgs field of the SM.
To obtain the list of CGCs implemented in \pyrate for the contraction of four arbitrary triplet fields of $\mathrm{SU}(2)$ (\rep{a}, \rep{b}, \rep{c}, \rep{d}) we can query the database 
\begin{Verbatim}[numbers=left,xleftmargin=20pt,formatcom=\color{cyan}]
Invariants SU2 [[2,True],[2],[2,True],[2]]
\end{Verbatim}
which\footnote{Note that \code{[2,True]} denotes the conjugated representation $\crep{2}$.} leads to three invariants, $\mathcal{C}^1,\ \mathcal{C}^2,\ \mathcal{C}^3$, out of which one vanishes ($\mathcal{C}^{3}$) once the substitutions $\rep{a},\rep{c}\rightarrow T^{\dagger}$ and $\rep{b},\rep{d}\rightarrow T$ have been performed. Working the details of the matching between these invariants and Eqs.~(\ref{eq:exsu2_1},\ref{eq:exsu2_2}) one arrives at the following relations for the CGCs of the $\lambda_{\Delta_2}$ term and $\lambda_{3}$ term, denoted $\mathcal{C}_{\lambda_{\Delta_2}}$ and $\mathcal{C}_{\lambda_3}$ respectively
\begin{equation}
	\mathcal{C}_{\lambda_{\Delta_2}}\rightarrow \mathcal{C}^1\, ,\, \mathcal{C}_{\lambda_{3}} \rightarrow \frac{1}{2}\mathcal{C}^1 + \frac{1}{\sqrt{3}}\mathcal{C}^2\, .
\end{equation}
This is input in the model file via the keyword {\it CGCs} as follows
\begin{Verbatim}[numbers=left,xleftmargin=20pt,formatcom=\color{cyan}]
QuarticTerms: {
'\lambda_{Delta_2}' : {Fields : [T*,T,T*,T], Norm : 1,CGCs: {SU2L: [1]}},
'\lambda_{3}' : {Fields : [[T*,T,T*,T],[T*,T,T*,T]], Norm : [1/2,1/sqrt(3)],
	CGCs: {SU2L: [1,2]}},
...
}
\end{Verbatim}
Similar relations are obtained for the other couplings\footnote{The details can be found at \url{http://pyrate.hepforge.org}.} and one can then generate the RGEs for all the quartic couplings as well as the gauge couplings. 

\begin{figure}[!h]
	\centering
	\includegraphics[width=\textwidth]{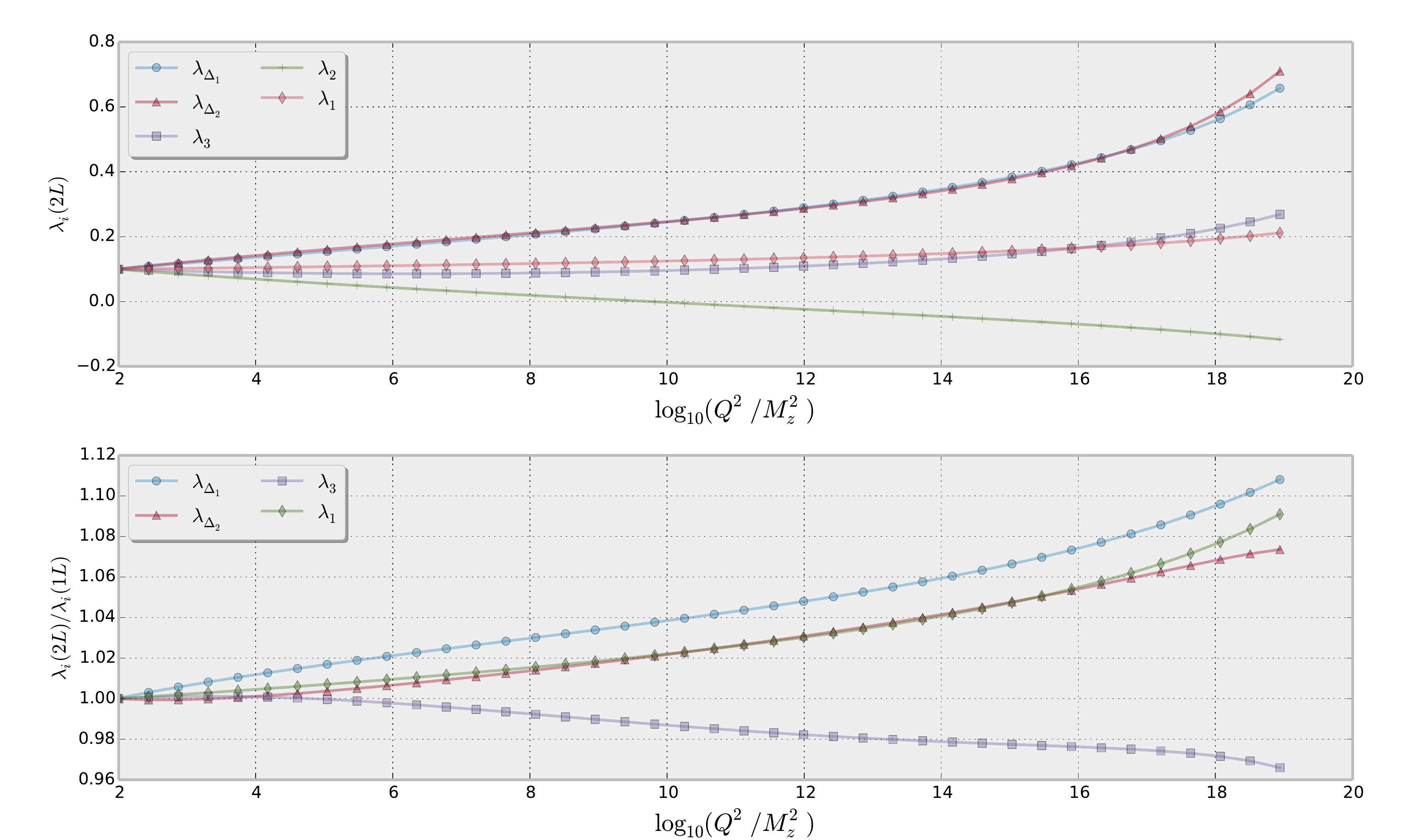}
	\caption{Evolution of the quartic couplings in a model where the SM has been supplemented by a complex $\mathrm{SU}(2)_L$-triplet scalar field. {\it Upper panel}: Quartic couplings at two-loop. {\it Lower panel}: Ratio of the two-loop over one-loop evolution; the effects are as large as 10\%.}
	\label{fig:explot1}
\end{figure}

Fig.~\ref{fig:explot1} upper panel shows the evolution with the energy scale of the various quartic couplings at two-loop\footnote{Note that this is not intended to be a phenomenologically viable example but simply an illustration of the capabilities of the \pyrate program.}. All the quartic couplings are set to 0.1 at $M_Z$ and then evolved according to the two-loop $\beta$-function. The lower panel of Fig.~\ref{fig:explot1} shows the ratio of the two-loop running over the one-loop running and illustrates the size of the corrections that one might expect from the two-loop $\beta$-functions; in this toy model, the modifications reach up to 10\% at the highest energy scales.

\section{Outlook}
\label{sec:outlook}

The above latest developments as well as the features described in this section have been combined into a new version, \pyratetwo\cite{lyonnet:2016pyr}. The main addition to this first version is the introduction of the kinetic mixing terms in the RGEs. Following~\cite{Fonseca:2013bua}, we have implemented all the effects of kinetic mixing at two-loop both in the gauge couplings and in the quartic, Yukawa, trilinear, fermion mass and scalar mass terms. This will constitute the first complete implementation of those terms. In addition, the RGEs for the fermion and scalar anomalous dimensions at two-loop have been added to the code. Finally, we wrote a new numerical output in \CC which significantly improves the speed to solve the coupled system of differential equation.

\section{Conclusion}

We have presented the \python program \pyrate that allows for an easy generation of two-loop RGEs for an arbitrary gauge field theory. With \pyrate, once the user has specified the gauge group and particle content of the model, the complete set of two-loop $\beta$-functions is derived. Numerical routines are automatically produced by \pyrate which allow for an easy and efficient way to solve the RGEs. The latest developments have been presented and in particular the new capability to deal with terms that have multiple gauge singlets. This was exemplified with the SM plus a complex triplet of $\mathrm{SU}(2)$ for which the complete set of two-loop $\beta$-functions has been obtained for the first time. Two-loop corrections in this toy model can reach up to 10\%. 

\Acknowledgments
I am grateful to Helena Kole\v{s}ov\'{a} and Ingo Schienbein for many useful discussions.

\addcontentsline{toc}{section}{References}
\bibliography{eprint_dpf2015.bbl}
\bibliographystyle{utphys}

\end{document}